\documentclass[conference]{IEEEtran}
\IEEEoverridecommandlockouts
\usepackage{cite}
\usepackage{amsmath,amssymb,amsfonts}
\usepackage{algorithmic}
\usepackage{graphicx}
\usepackage{textcomp}
\usepackage{booktabs}
\usepackage{xcolor}

\def\BibTeX{{\rm B\kern-.05em{\sc i\kern-.025em b}\kern-.08em
    T\kern-.1667em\lower.7ex\hbox{E}\kern-.125emX}}
\begin{document}

\title{DEEP FUNCTIONAL MULTIPLE INDEX MODELS WITH AN APPLICATION TO SER\\
}

\author{\IEEEauthorblockN{Saumard Matthieu}
\IEEEauthorblockA{\textit{VISION-AD Team, Labisen} \\
\textit{ISEN Yncr\'ea Ouest}\\
Brest, France \\
matthieu.saumard@isen-ouest.yncrea.fr}
\and
\IEEEauthorblockN{El Haj Abir}
\IEEEauthorblockA{\textit{VISION-AD Team, Labisen} \\
\textit{ISEN Yncr\'ea Ouest}\\
Caen, France \\
abir.el-haj@isen-ouest.yncrea.fr}
\and
\IEEEauthorblockN{Napoleon Thibault}
\IEEEauthorblockA{\textit{VISION-AD Team, Labisen} \\
\textit{ISEN Yncr\'ea Ouest}\\
Brest, France \\
thibault.napoleon@isen-ouest.yncrea.fr}
}

\maketitle

\begin{abstract}
Speech Emotion Recognition (SER) plays a crucial role in advancing human-computer interaction and speech processing capabilities. We introduce a novel deep-learning architecture designed specifically for the functional data model known as the multiple-index functional model. Our key innovation lies in integrating adaptive basis layers and an automated data transformation search within the deep learning framework. Simulations for this new model show good performances. This allows us to extract features tailored for chunk-level SER, based on Mel Frequency Cepstral Coefficients (MFCCs). We demonstrate the effectiveness of our approach on the benchmark IEMOCAP database, achieving good performance compared to existing methods. 
\end{abstract}


\section{Introduction}

Emotion recognition is an essential aspect of human-robot interaction, 
as it allows for a more natural and effective means of communication. 
Humans communicate their emotions through various modalities, 
including facial expressions, body language, and speech. However,
 among these modalities, speech plays a crucial role in emotion recognition, 
 as it is one of the most reliable and informative ways to convey emotional information. 
 Through speech, individuals can convey not only the content of their message but also 
 the underlying emotional state, including tone, pitch, and intonation. Moreover, 
 speech can also provide insights into the speaker's personality, cognitive state, and overall well-being.
  As such, developing accurate and efficient methods for speech emotion recognition is critical for creating more effective and responsive human-robot interfaces.


Support Vector Machines have been shown to be effective for speech emotion recognition. 
They are able to learn complex decision boundaries and can handle high-dimensional data, such as speech.
Other types of classifiers that have been used for speech emotion recognition include decision trees, k-nearest neighbors, and neural networks.
Speech emotion recognition has made rapid progress in recent years with the use of deep learning
and convolutional neural networks (CNN) \cite{Dossou2021}, transformer, attention and self-supervised learning \cite{Peng2021,Zhu2022,goncalves2022,Morais2022} methods. 

Since the pioneer monographs \cite{Ramsay_2005} and \cite{ferraty},
 functional data analysis (FDA) has become a vibrant
 field of research in the statistical community due to it vast applications in  various sciences, including  
astronomy, chemo-metrics, health and finance \cite{robbiano2016,saeys2008,cao2020}. The core objects of FDA are 
curves or functions of a separable Hilbert space like $L^2[0,1]$. However, 
statistical study of random variables of more general functions space like Banach
 space \cite{Bosq2002} or curves on manifold \cite{chen_2012} falls within the 
spectrum of FDA.  The fundamental frequency curve of an utterance has already been considered
 as a functional object \cite{arias13, Arias2014}. Recently, \cite{Tavakoli_2019} analysed dialect
sound variations across Great Britain using a spatial modeling approach that employs MFCC.
 In this article, we extend this point of view to speech emotion recognition.
Each coefficient of MFCC can be interpreted as a functional data variable. With a collection of coefficients, it is therefore natural to establish
 a correspondence between a speech recording and a multivariate functional data object. The number of covariates of the multivariate
functional data object is the number of coefficients of the MFCC used. A transformation of the multivariate functional object
is employed to serve as the final multivariate object, which is considered in a functional  multiple-index model.

In our work, we propose a novel approach for speech emotion recognition that involves treating Mel Frequency Cepstral Coefficients (MFCCs) as a functional data object.
 By doing so, we can represent each coefficient as a function of time, and thus extract  information from the speech signal.
  However, to compare functional data objects between samples with different duration, we need to preprocess the MFCC.
  This is achieved by splitting the MFCC in chunks,
   which allows us to represent each sample as a multivariate functional object. An abundant literature on chunk-level SER exists, see \cite{lin2020efficient},\cite{kumawat2021} and \cite{lin2023 } for example.
 This multivariate object can then be transformed into another multivariate 
 object using a suitable transformation, enabling us to use the functional multiple-index
  model for multivariate functional covariate to classify the emotions of the speaker.

This novel approach is particularly advantageous as it allows us to consider each MFCC as a functional variable, which captures 
the dynamic nature of speech and its relationship to emotions. By using a multivariate functional object,
 we can compare it across samples with different duration. 
 Moreover, the functional multiple-index model allows us to consider the interdependence between the different coefficients of the MFCC,
  providing a more accurate and comprehensive representation of the speech signal. Overall, our approach shows interesting perspectives for improving speech emotion recognition.

The paper is organized as follows.
We highlight the previous work in the following section.
Next, we describe the method in details with our contributions. We propose to evaluate our method on IEMOCAP database
described in the following fourth section along with the results and comparison with other methods. We conclude and discuss the results in the final section.


\section{Related work}

\subsection{On functional data models}

The functional single index models have been studied both from a theoretical and practical point of view in \cite{jiang2020,Jiang11,ferraty2011}. Some authors \cite{muller2008} and \cite{wong2019} use functional additive models that can be more stable than functional multiple-index models. The authors of \cite{rao2023} use deep neural network strategy to learn the parameter of the functional linear model by using a new functional neuron that can learn the functional representation with functional inputs. The article of \cite{yao21} proposes a novel neural network that learns the best basis functions for supervised learning tasks with functional inputs. They called it AdaFNN. The AdaFNN network parameterizes each basis node with a micro neural network that outputs a score of the input function $X(t)$,
which is the inner product between 
the basis function $\theta(t)$ and the input function:
\begin{equation*}
c=\langle \theta,X\rangle = \int \theta(t)X(t)\text{d}t.
\end{equation*}
They introduce neural networks that employ
a new Basis Layer whose hidden units are each
basis functions themselves implemented as a micro
neural network. On the link between deep learning and FDA, there exist other approach based on a functional layer developed by \cite{rossi2002} investigated in details by \cite{wang2019} and  \cite{rao2023}. We present in the next section our model and a simulation comparison with \cite{rao2023}.
The advantage of the approach in \cite{yao21} is that we can use all the deep learning tools without rewriting the back-propagation algorithm which is necessary in \cite{rao2023}.

Let us introduce the functional single index model in the regression context.
 Let $(Y_i,X_i)_{i=1,\ldots,n}$ be the set of data where $X_i$ represent the $i$-th functional data variable in $L^2[0,1]$, the space of square integrable function on $[0,1]$, and $Y_i\in\mathbb{R}$.
 \begin{equation}
 Y_i = g\left(\langle X_i,\theta\rangle\right)+\varepsilon_i,
 \end{equation}
where $\varepsilon_i$ are the error terms, $\langle . , . \rangle$ is the inner product in $L^2[0,1]$.
 $g$ and $\theta$ are unknown function to be estimated.
Generally, $g$ is estimated from a nonparametric framework with a kernel and $\theta$ is estimated from a basis decomposition or 
a functional principal component analysis. 
We propose to extend this definition to a functional multiple-index model for multivariate functional data covariate, see section 3.3.

\subsection{On speech emotion recognition}

There is a rich literature on speech emotion recognition using various deep learning architectures.
 We can cite \cite{kim22d,prabhu22,dhamyal22,perez22} which use self-attention, Bayesian neural networks and positional encoding. 
 The article of \cite{vaaras22} proposes to make silence representations of speech.
  The article of \cite{baruah22} makes progress by creating self-supervised learning for SER tasks.

\section{Method}

\begin{figure*}
\centering
\includegraphics[width=7in]{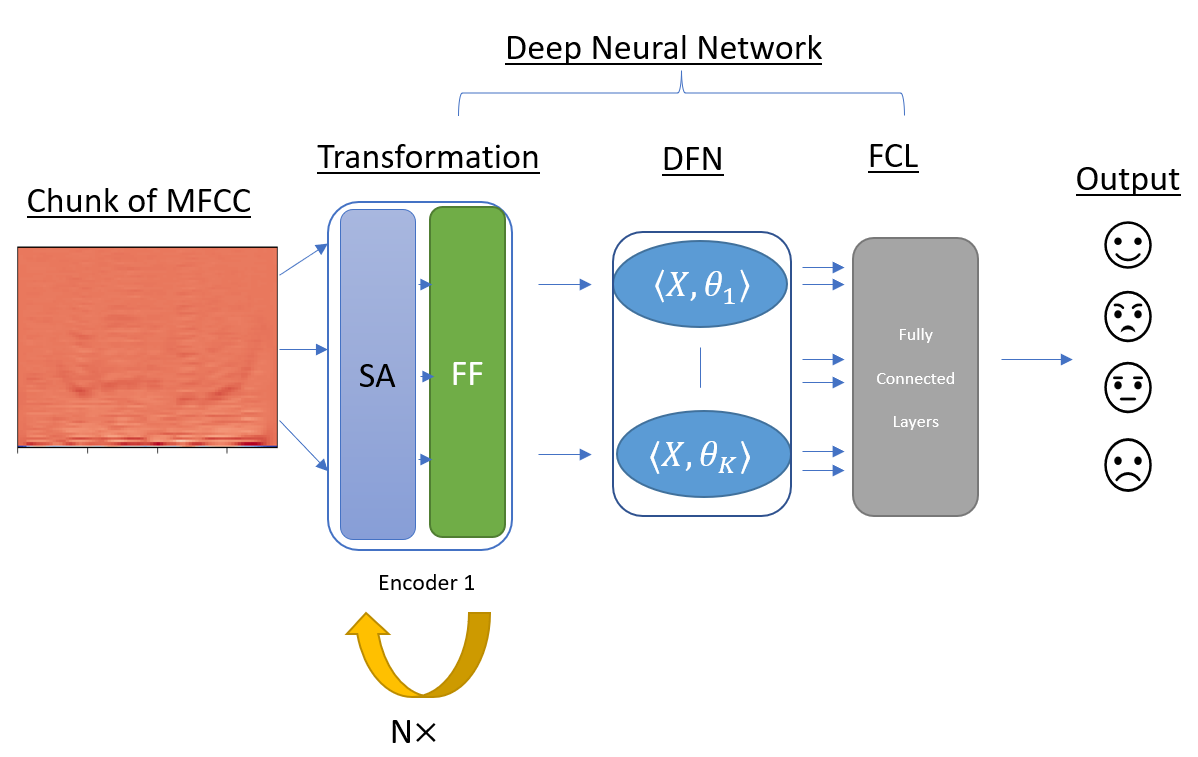}
\caption{Proposed method. SA: Self-Attention, FF: Feed-Forward network, DFN: Deep Functional Network, FCL: Fully Connected Layers.}
\label{figschema}
\end{figure*}

\subsection{MFCC}

Let us recall the method to calculate the MFCC.
With at hand a raw audio signal data representing by a time series $s(t)$ for $t=1,\ldots,T.$ 
We can consider that $s$ is defined for $t\in\mathbb{Z}$ by adding $0$ to non-value.
Let $w_M(t)$ for $t\in\mathbb{Z}$ be a window function of width $M$, we can define the spectrogram of the audio signal $s(t)$ by

\begin{equation}
\text{Spec}(t,\omega)=\vert \sum_{u=1}^T s(t-u) w_M(u) \exp (-i\omega u) \vert,
\end{equation}
for $t=1,\ldots,T,\omega\in[0,2\pi]$.

Hence, we can define the Mel spectrogram, which is a filtered version of the spectrogram to represent the human ear auditory system:

\begin{equation}
\text{MelSpec} (t,f)=\sum_{k=0}^{N-1}\text{Spec} \left(t,\frac{2k\pi}{N}\right)b_{f,k}
\end{equation}
for $f=0,\ldots, F$, with $b_{f,k}$ representing the set of the Mel-scale filter bank. Recall that the Mel scale is $m=2595 \log_{10}(1+\frac{f}{700})$.
The MFCC are then:
\begin{equation}
\text{MFCC}(t,m)=\frac{1}{F}\sum_{f=0}^F \log \left( \text{MelSpec(t,f)}\right) \exp (i(2\pi\frac{m-1}{F+1})f ),
\end{equation}
for $m=1,\ldots,n_{MFCC}$.


Note that there exist variations of the definition of MFCC in the literature. For example, we can use a Discrete Cosinus Transform (DCT) to return to the time scale.

 \subsection{Functional multiple index models}

For extracting new features based on MFCC, we employ a deep functional multiple index model.
 Let $(Y_i,X_i)_{i=1,\ldots,n}$ be the set of data where $X_i$ represent the $i$-th multivariate functional object and $Y_i$ is its associated label. 
Let us introduce the model.
\begin{align}
Z_i =& T(X_i)\\
Y_i =&g\left( \langle Z_i^1,\theta_1\rangle,\ldots, \langle Z_i^p,\theta_K\rangle\right)+\varepsilon_i,
\end{align}
where $T: (L^2[0,1])^{p}\rightarrow (L^2[0,1])^{p}$ is a transformation of the multivariate functional data,
 $\theta_j$ are the indexes (functions of $L^2[0,1]$), $g: \mathbb{R}^{p\times K} \rightarrow \{0,\ldots, C-1\}$ is the link function, $C$
 is the number of classes, $\langle . , . \rangle$ is the inner product of $L^2[0,1]$ functions and the power $j$ of the variable $Z_i$ is the
 $j$-th component function of $Z_i$ as $Z_i$ is a multivariate functional data.
 
 The unknown functions are the transformation $T$, the link function $g$ and the indexes $\theta_j$, $j=1,\ldots,p\times K.$ The transformation $T$ is inferred by a transformer encoder architecture. Next, we apply an AdaFNN network to each functional variable, the output represents new features extracted from the MFCCs. Thus, the $\theta_j$ are estimated by a deep functional network (DFN) based on a concatenation of the AdaFNN module. Then, we apply a fully connected layer to estimate the link function $g$.

\subsection{Deep neural network of the model}

With a speech recording, we calculate the MFCC associated and cut it with an overlapping method. Then, we can feed the Deep Neural Network with this chunk of MFFC
seen as a multivariate functional object. First, we make a transformation of the MFCC chunk. There is undoubtedly 
a link between the size of MFCC and the information about the emotion of the speech. But we do not know how is this link. So we apply a transformation $T$
on the resized MFCC to accurately predict the emotion contained in the speech. The transformation module is made of a stack of $N$ transformer encoders with self-attention and Feed-Forward network. 
  The second module is the neural network of the paper \cite{yao21} which is generalized to adapt to the multivariate context. 
  In few words, the paper of \cite{yao21} proposes to integrate the data by an adaptive function by a numerical scheme . So, the outputs of this second module are simply
  the $L^2$ product between each component of the transformed MFCC and an adaptive function. The third module of our network is a classical fully connected layer.
The proposed method is represented globally in Figure \ref{figschema}.

The three modules of our new network architecture are: \begin{enumerate}
\item The transformation module made of $N$ transformer encoders.
\item The Deep Functional Network that outputs the scores of multivariate functional variable.
\item A Fully Connected Layers to classify the emotions.
\end{enumerate}

\subsection{Simulations}

In order to show the ability of our proposed method, we study three different scenarios. For completing the study on other potential models using adaptive layer, we refer to the simulation section of \cite{yao21}.
We simulate four functional covariates $X^{(j)}\, j=1,2,3,4$ coming from four different processes: exponential variogram ($j=1$), Brownian ($j=2$), Fractional Brownian ($j=3$) and Gaussian process with Mat\'ern covariance function. These curves are evaluated at $30$ equally-spaced time points from $[0; 1]$. The unknown parameter functions are $\beta_1(t)=5\sin (2\pi t)$, $\beta_2(t)=5\sin (3\pi t)$, $\beta_3(t)=3\cos (2\pi t)$ and $\beta_4(t)=3\cos(3\pi t)$. A Gaussian error of variance $0.04$ has been added to the regression term.
Let us introduce the three different scenarios:
\begin{equation*}
(S1)\,    Y_i = g\left(\sum_{j=1}^4 \langle \beta_1, X_i^{(j)}\rangle,\sum_{j=1}^4 \langle \beta_2, X_i^{(j)}\rangle\right) +\varepsilon_i 
\end{equation*}
with
\begin{align*}
    g(a,b)=a^2+b^2.
\end{align*}

\begin{equation*}
 (S2)\,   Y_i = g\left(\sum_{j=1}^4 \langle \beta_1, X_i^{(j)}\rangle,\cdots,\sum_{j=1}^4 \langle \beta_4, X_i^{(j)}\rangle\right) +\varepsilon_i 
\end{equation*}
with
\begin{align*}
    g(a,b)=a^2+b^2+c^2+d^2.
\end{align*}

\begin{equation*}
\begin{split}
  (S3)\,  Y_i =& (\langle \beta_1, \Tilde{X_i}^{(1)}\rangle+\langle \beta_2, \Tilde{X_i}^{(2)}\rangle+ \\\langle \beta_3, X_i^{(3)}\rangle\times &\langle \beta_4, X_i^{(4)}\rangle)^2+ \\ (\langle \beta_1, \Tilde{X_i}^{(1)}\rangle\times&\langle \beta_2, \Tilde{X_i}^{(2)}+\langle \beta_3, X_i^{(3)}\rangle+\langle \beta_4, X_i^{(4)}\rangle)^2 +\varepsilon_i 
\end{split}
\end{equation*}

\begin{table}[th]
  \caption{Performance on simulations. We report RMSE.}
  \label{tab:simu}
  \centering
  \begin{tabular}{|c|c|}
    \toprule
   Scenario & RMSE\\
    \midrule
 $S(1)$ & $0.085$\\
 $S(2)$ & $0.074$ \\
 $S(3)$ & $0.031$ \\
    \bottomrule
  \end{tabular}
  
\end{table}

To make an evaluation of the transformation part, we change $X^{(1)}$ by $\Tilde{X}^{(1)}=(X^{(1)})^2$ and take $\Tilde{X}^{(2)}=| X^{(2)}|$ in the last scenario. We do not have the code of the paper \cite{rao2023}, so we could not compare to their method. In table \ref{tab:simu}, the simulations reflect a good behaviour of our approach, even with complex behaviour and transformations in variables. 

\section{Application to SER}

\subsection{Database}

The dataset IEMOCAP \cite{iemocap} contains
approximately 12 hours of speech from 10 speakers. The literature
selects a total of 5531 utterances that are labeled with one of
five categories: Happy, Angry, Neutral, Sad, Exited. This set is
reduced to four categories by merging Exited and Happy
into a single category. 

For evaluating our method, we use the protocol designed in \cite{antoniou2023designing}. Namely, we perform a $10$-fold Speaker Independent cross-validation with one speaker as test, eight as training and one as validation set.

\subsection{Implementation and hyperparameters}

The code is written in python and use the pytorch library including its modules. We use the Adam optimizer with a learning rate of $3\times 10^{-4}$ and a focal loss with the $L^2$ penalty on the basis parameter.
 We set the batch size to $32$ with $15$ epoch and return the best model on the validation set. 
 We do not tune the number of basis $K$, we use the best result of \cite{yao21}, namely $4$ basis functions.
 We set the number of MFCC $p$ to $40$. And, we make chunks of the MFCC with a duration of $64$.
 So, finally, the size of the final MFCC is $(64,40)$. We use $N=2$ transformer encoder layers.
 We use in the DFN of the basis layers an hidden FF of three connected layers of $128$ each.
 The subsequent FCL network is two dense layer with an $\tanh$ activation function, dropout of $0.2$, follow up by a projection to the $4$ classes.

\subsection{Results}

We calculate two metrics namely WA for weighted accuracy (overall accuracy)
 and UA for unweigthed accuracy (average of the recall). We compare with \cite{shin2023,Peng2021}
 which are the best results on IEMOCAP with four emotions and speech only, see Table~\ref{tab:perf}.
 It is worth mentioning that \cite{shin2023} do not only use MFCC but also Zero-crossing rate (ZCR), root mean square (RMS), Mel vector, chroma. And \cite{Peng2021} use MFCC and first and second order frame-to-frame difference.
\begin{table}[th]
  \caption{Performance of different approaches on the IEMOCAP testing set. We report WA and UA.}
  \label{tab:perf}
  \centering
  \begin{tabular}{ c|c|c }
    \toprule
   Method & WA& UA\\
    \midrule
 GRU \cite{shin2023} &$ 64.95$&$.$ \\
 AUDIO-CNN \cite{Peng2021} &$66.6$&$68.4$\\
 DFN (ours) & \textbf{57.27} &\textbf{56.43}\\
    \bottomrule
  \end{tabular}
  
\end{table}

The results cannot be compared directly because we make our analysis on chunk-level. We think that the performance of our approach with the protocol of \cite{antoniou2023designing} can be improve by considering all the chunks of the audio record considered.

\section{Discussion}

In this method, we choose to make chunks with an overlap of $25\%$ of the duration of the chunk. Recently, \cite{lin2020efficient} and \cite{lin2023} choose dynamically the percentage of overlapping with a significant improvement in accuracy.
We can enhance our method by dynamically choosing the chunk overlapping by using the method in \cite{lin2020efficient} and \cite{lin2023}. Moreover, the features can be passed to a recurrent neural network (RNN) (like LSTM, Bi-LSTM, GRU) to make a global decision on the whole audio input. Adding a RNN on top of the model may enhance the results, and that can be done in a two-stage or end-to-end learning.

\section{Conclusion}
In conclusion, the article presents a promising advancement in SER using a novel model. The results, validated through simulations and tested on the IEMOCAP dataset at the chunk-level, demonstrate satisfactory performance. The model leverages new features extracted from MFCC and relies on functional data, showcasing an innovative approach to emotion detection in speech. This development not only contributes to the ongoing evolution of SER methodologies but also highlights the potential for further exploration of functional data in enhancing emotion recognition systems.

\bibliographystyle{IEEEtran}
\bibliography{emotions}

\end{document}